\documentclass[11pt,twoside]{article}
\usepackage{asp2004}
\usepackage{psfig}
\usepackage{epsf}
\usepackage{graphics}
\usepackage{lscape}
\def\edcomment#1{\iffalse\marginpar{\raggedright\sl#1\/}\else\relax\fi}
\reversemarginpar

\begin{document}
\title{Redshifts and Distribution of ACO Clusters of Galaxies}
\author{Heinz Andernach}
\affil{Departamento de Astronom\'{\i}a, Universidad de Guanajuato, \\
Apartado Postal 144, 36000 Guanajuato, Gto, Mexico}
\author{Erik Tago, Maret Einasto, Jaan Einasto}
\affil{Tartu Observatory, EE--2444 T\~oravere, Estonia}
\author{Jaak Jaaniste}
\affil{Institute of Physics, Estonian Agricultural University, \\
Kreutzwaldi 64, EE-51014 Tartu, Estonia}

\begin{abstract}
The June~2004 version of our compilation of measured redshifts for 
clusters in the Abell-ACO catalogue lists redshifts for 3715 
clusters/subclusters in 3033 distinct 
(2396 A- and 637 S-) clusters, 67\,\% of these with $N_z\ge\,3$ 
galaxies measured. We provide velocity dispersions ($\sigma_V$) 
for 1875 (sub)clusters towards 1353 unique ACO clusters.
The median $\sigma_V$ is 650\,km\,s$^{-1}$ for A-(sub)clusters and 575\,km\,s$^{-1}$ for
S-(sub)clusters, and $\sigma_V$ clearly increases with both, 
$N_z$ and richness, and also, somewhat surprising, with later Bautz-Morgan 
type of the clusters. We show examples of supercluster properties based on these data.
\end{abstract}

\section{Introduction, Methodology and Problems encountered}

Our compilation of measured redshifts for clusters from the catalogue
of Abell, Corwin, \& Olowin (1989, ACO) has been maintained since 1989
(Andernach 1991; Andernach, Tago, \& Stengler-Larrea 1995 (ATS95);
Andernach \& Tago 1998 (AT98)) and is the only such compilation
including both A- and S-clusters.  Different from previous compilations
(e.g.\ Struble \& Rood 1999, SR99, who include A-clusters only), we
systematically scan the literature for galaxy redshifts.  Surveys like
the LCRS (Shectman et al.\ 1996), 2dF, 6dF (Jones et al.\ 2004), and
SDSS (Abazajian et al.\ 2004) prove to be rich in ACO cluster redshifts.
Our selection criteria for galaxy redshifts, and the cluster parameters
we include in our compilation can be found in ATS95 and AT98.  

%

The number of references contributing cluster redshifts has
increased so much that one of us (HA) now compiles all individual
galaxy positions and redshifts, so as to ease the merging of different data
sources.  For clusters with large $N_{z}$ and {\it several} papers to 
merge, this work is in progress.  Our current compilation is based 
on 685 references, growing by $\sim$40 per year.
Sixty percent of the 3715 mean redshifts, and 86\,\% of the 1875 
velocity dispersions are based on our own calculations, using individual
galaxy data. Over 330 redshifts are based on the merging of 
galaxy data from four or more references.  We do not include any 
photometric redshifts nor any galaxy velocities with errors
$\ga$600\,km\,s$^{-1}$.

In order to merge galaxy data from different sources for the same cluster,
we need individual galaxy positions, velocities, and errors.  Lack of
these data prevents the merging of many data sources or calculation of $\sigma_V$
(e.g., we list over a dozen clusters with $N_{z}\ge10$ but no $\sigma_V$ reported). 
Often publications do not state whether redshifts are 
geo-, helio- or galactocentric.
The fact that the data releases of large projects like 2dF, 6dF, SDSS,
etc., tend to be {\it cumulative} (i.e.\ each release contains previous
ones, often with reprocessed data) rather than {\it incremental}, makes
the updating of our compilation very tedious.

To calculate $\overline{z}$ and $\sigma_V$ for a cluster, we first search
for any relative maxima in the redshift distribution of galaxies within
the cluster area. Around each relative maximum we include into a single
cluster component all galaxies within $\pm 2500$\,km\,s$^{-1}$ from that maximum.
We use this value also as a minimum velocity gap for subclusters
closely spaced along the line of sight. Only if subclusters were 
published with a smaller separation in velocity (mostly due to 
components separated in the plane of the sky), we adopt them from 
the literature {\it as is}.

Apart from three known pairs of duplicate clusters in ACO, and six further
pairs reported by AT98, we propose A3742=S\,924 as an additional identity.

\section{Current Status}
Redshifts (based on at least one galaxy measured) are now available for
59\,\% of all 4076 A-clusters and for 54\,\% of all 1174 supplementary
southern S-clusters, which is an almost 4-fold increase over Abell et al.\ (1989).
Only 2.7\,\% of the redshifts are beyond a factor two from their photometric 
estimates (see Sect.\,3).
A significant improvement over previous compilations is that 
67\,\% (compared to 43\,\% in SR99) of the redshifts are based on 
$N_z\ge\,3$ measured galaxies, and can thus be considered ``reliable''.
This, as well as the almost 3-fold increase in the number of known
velocity dispersions over SR99, is due to our efforts to merge all 
available data sources, especially in the regime of low $N_{z}$.

There are 1245 (sub)clusters with $N_{z}>$10, of which 276 have $N_{z}>$50
(the typical minimum for dynamical studies), and 95 (sub)clusters have
$N_{z}\ge100$. The total number of galaxies involved for all our listed
(sub)clusters is $\sim$56,800 (9500 of which in S-clusters), including
some overhead for overlapping clusters for which it is not possible to
assign galaxies uniquely to one cluster.

We quote $\sigma_V$ for 1353 different Abell clusters (1080 A- and 273
S-clusters, for a total of 1875 subclusters). The median $\sigma_V$ for
all 1875 (sub)clusters is 636\,km\,s$^{-1}$, and their distribution 
shows an almost Gaussian main peak at $\sim$630\,km\,s$^{-1}$ with
a dispersion of $\sim$275\,km\,s$^{-1}$, followed by a weak tail out to 
2000\,km\,s$^{-1}$ (probably containing some line-of-sight superpositions or mergers).

\section{Some Exploratory Analysis}
 
For the rich ACO clusters (A-names) we use the redshift estimate by
Peacock \& West (1992, kindly provided by M.\,West), while for the
S-clusters we use a function originally proposed by Abell et al.\
(1989) for southern A-clusters, but scaled down by 30\% on the basis
of 196 S-cluster redshifts (Andernach 1991).  Now, with three times the
number of S-cluster redshifts in hand, we confirm S-clusters to have on average
30\% lower $z$ than the ACO estimate. It is not clear whether this is due
to an excess foreground contamination of redshifts, or due to a
systematic overestimate of $m_{10}S$ of the clusters caused by background
contamination.  We do {\bf not} confirm the claim by AT98 that S-clusters
suffer more from line-of-sight superposition than A-clusters: currently
16\,\% of both A- and S-clusters appear with more than one redshift entry,
and it rather seems that deeper and deeper surveys (like 2dF or SDSS)
tend to find more clusters along the line of sight.

Redshifts are now available for all except 60 (1.5\%) rich (A-) clusters 
with estimated redshift z$_{est}\la$0.1, while 17 S-clusters (7\,\%)
with $z_{est}<$0.05 (and 240, or 32\,\%, with $z_{est}<$0.1)
still lack a measured redshift. 
There is still plenty of high-galactic-latitude sky with galaxies 
brighter than $15\fm5$ unexplored in redshift. However,
these gaps are getting smaller: e.g., we found that only the increment of 6dF-FDR
(Jones et al.\ 2004) over 6dF-EDR yields new redshifts for $\sim$250 
clusters with no previous $z$, about half of these with $z_{est}<0.1$.
Based on now 15 years of compilation experience we anticipate that 
only in $\sim$10 years from now the redshift 
completeness of the ACO catalogue may get close to 100\,\%.

The distribution of $\sigma_V$ for the 1875 (sub)clusters
depends strongly on $N_z$.  We find median values of $\sigma_V$ of 
556\,km\,s$^{-1}$ for 651 (sub)clusters with $N_z<10$,
616\,km\,s$^{-1}$ for 564 (sub)clusters with $10\leq N_z \leq20$, and
703\,km\,s$^{-1}$ for 597 (sub)clusters with $N_z>20$. These medians
are slightly lower than those reported in AT98 due to a more careful
clipping of outliers.  As shown by Plionis et al.\ (2004)
$\sigma_V$ increases with both $N_z$ and with Abell count $N_{Ab}$.
This correlation may be partly due to an observer's 
tendency to measure more redshifts in richer clusters and thus the 
$\sigma_V- N_z$ correlation may be a reflection of the known richness 
dependence of $\sigma_V$.

The morphological classification available for most ACO clusters is 
the ``Bautz-Morgan'' (BM) type. Using only subclusters with the 
highest $N_z$, we find that the mean and median $\sigma_V$ for 664 
(sub)clusters of ``early'' BM\,type ($\leq$II) 
are {\it lower} by $\sim$45\,km\,s$^{-1}$ than for 614 (sub)clusters of 
``late'' BM\,type ($>$II). \linebreak[4]
A Kolmogorov-Smirnov test gives only a 1.8\,\% chance for 
these samples to be drawn from the same population.
The direction of this trend is surprising, as earlier BM types are expected 
to be more evolved and to have higher X-ray luminosity, and thus to show a higher
$\sigma_V$ (David, Forman, \& Jones 1999). These authors found that 
X-ray luminosity is correlated with both Abell richness class $R$ and BM type,
but also that $R$ may be overestimated in the southern ACO. Using
updated X-ray samples and more cluster parameters, like $\sigma_{V}$, we plan
to shed more light on the relation between cluster dynamics and BM type.

\section{Applications: Large-Scale Structure and Superclusters}
  
We used our compilation in the past to establish  
the presently most complete catalogues of superclusters (SCL) of 
Abell- and X-ray clusters (e.g.\ Einasto et al.\ 2001),
and showed that X-ray clusters are more strongly clustered into 
superclusters than are optical clusters.  We used the SCL catalogue 
to confirm that the richest superclusters occupy a more or less 
regular lattice of 120\,$h_{100}^{-1}$\,Mpc grid size 
(Einasto, et al.\ 1994, Saar et al.\ 2002). 
Based on the {\it current} compilation we extended our SCL 
catalogue out to $z=0.15$. 
In particular, several very interesting SCLs were revealed.

The most prominent Abell supercluster crossed by the Northern LCRS 
slices is  SCL\,126 ($z$=0.084) in the direction of Virgo.
Four out of a total of seven ACO member clusters of this SCL
lie within a sphere of diameter $\sim10\,h_{100}^{-1}$\,Mpc.  
Three clusters in this SCL are strong  X-ray sources, and
four contain radio sources. This
makes SCL\,126 one of the most unusual superclusters currently known.  
The shape ellipsoids, based on Abell clusters, 
on LCRS Loose Groups (LCLG, Tucker et al.\ 2000),
and on LCRS galaxies in SCL\,126, all have axis ratios of 
$\sim$1:4, and are located perpendicular to the line of sight
(Einasto, M. et al.\ 2003b). 
This may be evidence for a ``squashing effect'' of galaxies falling 
into SCLs (Kaiser 1987), accompanied by merging and other 
processes causing X-ray and radio emission from the clusters.
The core of SCL\,126 may have started to collapse (Gramann \& Suhhonenko 2002).

Another prominent SCL is ``Horologium--Reticulum'' (SCL\,48 at $z\sim$0.064),
crossed by all southern LCRS slices.  This supercluster consists of 
several concentrations of Abell clusters and LCLGs which are connected 
by filaments of galaxies, groups and clusters that surround underdense regions 
(Einasto, M. et al.\ 2003b; Einasto, J. et al.\ 2003b).

Galaxy groups from the LCRS and SDSS in high-density environments \linebreak[4]
(superclusters of Abell clusters) 
are also richer and more massive than groups in low-density environments 
(Einasto, M. et al.\ 2003a,\,b; Einasto, J. et al.\ 2003).

These results indicate that superclusters, as high-density environments, 
play a major role in the formation and evolution of galaxy systems.

\acknowledgments 
HA wishes to thank Mike Read (ROE) for providing 6dF-EDR and 6dF-DR1
in ASCII format. HA acknowledges financial support from CONACyT grant 
40094-F and from the conference organizers.

\end{document}